\documentclass{elsart}
\usepackage{amssymb}
\usepackage{graphicx}
\usepackage{epsfig}
\begin{document}
\begin{frontmatter}
\title{\Large \bf  Diffusion Anomaly \\ in a three dimensional lattice gas}
 \author[ufpampa]{Mauricio Girardi}
  \ead{girardi@fisica.ufsc.br}
\address[ufpampa]{Universidade Federal de Pelotas - UNIPAMPA/Bag\'e,
Rua Carlos Barbosa SN, CEP 96400-970, Bag\'e, RS, Brasil.}
\author[ufrgs]{Marcia Szortyka}
\ead{szortyka@if.ufrgs.br}
 \author[ufrgs]{Marcia C. Barbosa}
\ead{marcia.barbosa@ufrgs.br}
  \ead[url]{ http://www.if.ufrgs.br/$\sim$barbosa}
 \address[ufrgs]{Instituto de F\'{\i}sica, Universidade Federal do Rio
  Grande do Sul \break Caixa Postal 15051, 91501-970, Porto Alegre, RS, Brazil.}
\begin{abstract}

We investigate the relation
between thermodynamic and dynamic properties
of an associating lattice gas (ALG) model. The ALG combines a three dimensional
lattice
gas with particles interacting through a soft core potential and orientational
degrees of freedom. From the competition between the directional
attractive forces and the soft core
potential results two liquid phases, double criticality and density anomaly.
We study the mobility of the molecules in this model by
calculating    the diffusion constant at a constant temperature, $D$.
We show that $D$
has a maximum at a density $\rho_{max}$ and a minimum at a density
$\rho_{min}<\rho_{max}$. Between these densities the diffusivity
differs from the one expected for normal liquids. We also show that in
the pressure-temperature phase-diagram
the line of extrema in diffusivity is close to the liquid-liquid critical point and it
is partially inside the temperature of maximum density (TMD) line.

\end{abstract}

\maketitle
  \end{frontmatter}

\section{\label{sec1}Introduction}

\begin{figure}[htb]
\begin{center}
\includegraphics[clip, angle=-90,scale=0.3]{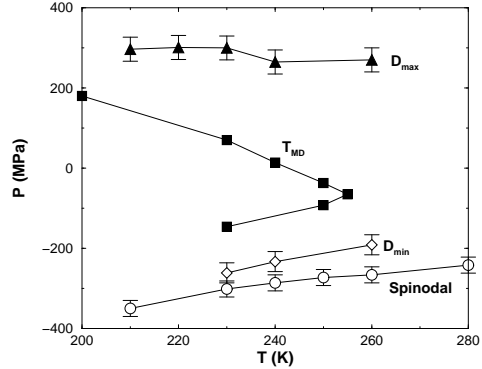}
\end{center}
\caption{Simulation data for SPC/E water from the work
of Netz \emph{et al.}.\cite{Ne01} The triangles determine
the loci where the diffusion has a local maximum value with
increasing density at fixed temperature, and the diamonds
mark its local minima. The squares
determine the temperature of maximum density line,
where density anomaly occurs, and the circles 
locate the liquid-gas spinodal.
 \label{cap:spce}}
 \end{figure}

\begin{figure}[htb]
\begin{center}
\includegraphics[clip=true,scale=0.4]{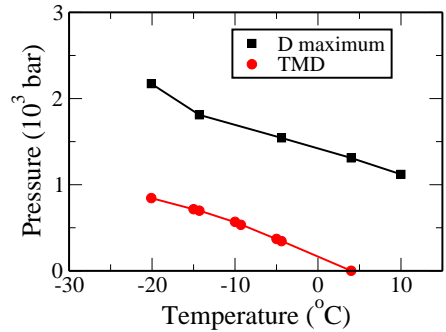}
\end{center}
\caption{Experimental data for water extracted from the work
of Angell \emph{et al.}.\cite{An76} The squares determine 
the loci where the diffusion has a maximum value with
increasing pressure at fixed temperature. The circles
stands for the temperature of maximum density (TMD) line, 
location where density anomaly occurs.
 \label{cap:experimental}}
\end{figure}

Most liquids contract upon cooling. This is not the case of water, a liquid
where the specific volume at ambient pressure starts to increase 
when cooled below $T=4 ^oC$ at atmospheric pressure \cite{Wa64}. This effect is called
density anomaly. Besides the density  anomaly, there are
sixty-two other anomalies
known for water.\cite{URL} The diffusivity  is one
of them. For normal liquids the diffusion coefficient,
D, decreases under compression. However, experimental results have show that for water at
temperatures approximately below 10$^o$C, the diffusion
coefficient increases under compression and has a
maximum (square symbols in Fig. \ref{cap:spce}).
For temperatures above 10 $^o$C,  D behaves as in
a normal liquid.
The temperature of maximum density (TMD) line
(circles in Fig. \ref{cap:experimental}),
inside which the density anomaly occurs, and
the line of maximum in diffusivity are located
in the same region of the pressure-temperature (P-T)
phase diagram of water.
\cite{An76} Simulations
also show thermodynamic and dynamic anomalies.
The simple point charged/extended (SPC/E) model for water
exhibits in the P-T phase diagram a TMD line.
The  diffusion coefficient has a maximum and a minimum
that define two lines at the P-T phase diagram,
the lines of maximum and minimum in the
diffusivity coefficient.
\cite{Ne01,Er01,Mi06a}
The TMD and the lines of maximum and minimum
in the diffusion are located at the same
region at the P-T phase diagram
for the SPC/E model (see Fig. \ref{cap:spce}).
Errington and Debenedetti \cite{Er01} and Netz \emph{et al}.\cite{Ne01}
found, in SPC/E water,
that there exists a hierarchy between the density and diffusion anomalies as follows.
The diffusion anomaly region, inside which the
mobility of particles grow
as the density is increased, englobes the density anomaly region,
inside which the system
expands upon cooling at constant pressure(see Fig. \ref{cap:spce}) .
Experiments for real water support these simulational results
(see Fig. \ref{cap:experimental}). \cite{An76}

It was proposed a few years ago that these anomalies are related to
a second critical point between two liquid phases, a low density liquid
(LDL) and a high density liquid (HDL) \cite{Po92}. This critical point
was discovered by computer simulations. This work suggests that the
critical point is located at the supercooled region beyond  the line of
homogeneous nucleation and thus cannot be experimentally measured.
In spite of  this limitation,  this hypothesis has been supported by
indirect experimental results \cite{Mi98,angell}.

One question that arises in this context is what kind of potential would be appropriated for describing the
tetrahedrally bonded molecular liquids, capturing the presence of
thermodynamic anomalies? Realistic simulations of water
\cite{St74,Be87,Jo00} have achieved a good accuracy in describing the
thermodynamic and dynamic anomalies of water. However, due to the high
number of microscopic details taken into account in these models, it
becomes difficult to discriminate what is essential to explain the
anomalies. On the other extreme, a number of isotropic models were
proposed as the simplest framework to understand the physics of
the liquid-liquid phase transition and liquid state anomalies. From the
desire of constructing a simple two-body isotropic potential capable
of describing the complicated behavior present in water-like molecules,
a number of models in which single component systems of particles
interact via core-softened (CS) potentials have been proposed.
They possess a repulsive core that exhibits a region of
softening where the slope changes dramatically. This region can
be a shoulder or a ramp 
\cite{St98,Sc00,Sc01,Fr01,Bu02,Bu03,Fr02,Ba04,Ol05,Ol06,He05a,He05b,Sk04,Ma04,He70,Ja98,Wi02,Ku05,Ca03,Ca05}.
Unfortunately, these models, even when successful in showing density
anomaly and two liquid phases, fail in providing the connection between
the isotropic effective potential and the realistic potential of water. 

It would, therefore, be desirable to have a theoretical framework which
retains the simplicity of the core-softened potentials but accommodates
the tetrahedral structure and the role played by the hydrogen bonds
present in water. A number of lattice models in which the tetrahedral
structure and the hydrogen bonds are present have been studied
\cite{Be72,Bes94,Ro96,Ro97,Ro98,Sa93,Sa96,Pr04,Pr06,Gi02,Gi04,Gi06b}. One of them, is the
three-dimensional model proposed by Roberts and Debenedetti
\cite{Ro96,Ro97,Ro98} and further studied by Pretti and Buzano
\cite{Pr04} defined on the body centered cubic lattice.
According to their approach, the energy between two bonded molecules rises
when a third particle is introduced on a site neighbor to the bond.
Using a cluster mean-field approximation and computer simulations
they were able to find the density anomaly and two liquid phases.  In this case,
the  coexistence between two liquid phases  may arise from the competition
between occupational and Potts variables introduced through a dependency of
bond strength on local density states.

Recently we have proposed an associating
lattice-gas (ALG) model which
retains the simplicity of the core-softened potentials but accommodates
the tetrahedral structure and the role played by the hydrogen bonds
present in water. This model system
is  a lattice gas with  ice variables \cite{Be33} which allows for a
low density ordered structure. Competition between the filling up of the
lattice and the formation of an open four-bonded orientational
structure is naturally introduced in terms of the ice bonding variables, and
no \emph{ad hoc} introduction of density or bond strength variations is
needed. In that sense, our approach bares some resemblance to that of
continuous softened-core models \cite{Si98,Tr99,Tr02}. Studying  this simple
model in two and three dimensions we were  able to find two liquid phases, two critical points and the
density anomaly \cite{He05a}\cite{He05b}\cite{Gi07}\cite{Ba07}.

In this paper, in the framework of the ALG, we address two
questions: (i) is the presence of
diffusion anomaly
related to the presence of density anomaly? (ii) if so, what
is the hierarchy between the two anomalies and the presence of a
second critical point?
 We show that the two anomalies are located in the same region
of the $P$-$T$   phase diagram, close to the second critical
point and that  the region on the $P$-$T$ phase diagram in which
the density anomaly is present
encloses the region in which the  diffusion anomaly exists.

In sec. II the
model is introduced and the simulation details are given. Sec. III is
devoted to the main results and conclusion ends this session.

\begin{figure}[htb]
\begin{center}
\includegraphics[clip=true,scale=0.4]{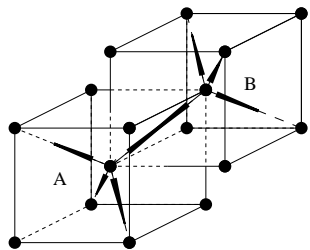}
\end{center}
\caption{The model.
 \label{cap:model}}
\end{figure}

\section{The Model}

Recently \cite{Gi07}, we have considered  a body-centered cubic lattice with $V$ sites, where each site
can be either empty or filled by a water molecule. Associated to each
site there are two kinds of variables: an occupational variables, $n_{i}$,
and an orientational one, $\tau_{i}^{ij}$. For $n_{i}=0$ the $i$ site
is empty, and $n_{i}=1$ represents an occupied site. The orientational
state of particle $i$ is defined by the configuration of its  bonding and 
non-bonding arms, as illustrated in Fig. (\ref{cap:model}). Four of them are the usual ice
bonding arms with $\tau_{i}^{ij}=1$ distributed in a tetrahedral arrangement,
and four additional arms are taken as inert or non-bonding ($\tau_{i}^{ij}$=0).
Therefore, each molecule can be in one of two possible states $A$ and $B$ as
illustrated in Fig. (\ref{cap:model}). A potential energy $\varepsilon$ is associated to
any pair of occupied nearest-neighbor ($NN$) sites, mimicking the van der
Waals potential. Here, water molecules have four indistinguishable arms that
can form hydrogen-bonds (HB). An HB  is formed  when two arms of $NN$ molecules
are pointing to each other with $\tau_i^{ij}=1$. An energy $\gamma$ is assigned
to each formed HB. 

In resume the total energy of the system is given by:
\begin{equation}
E=\sum_{(i,j)}n_{i}n_{j}\left(\varepsilon
+\gamma\tau_{i}^{ij}\tau_{j}^{ji}\right)\,\,.
\label{en}
\end{equation}

The interaction parameters were chosen to be $\varepsilon>0$ and $\gamma<0$,
which implies in an energetic penalty on neighbors that do not form $HBs$. From
this condition results the presence of two liquid phases and the density anomaly.

For studying the mobility, we have performed
 Monte Carlo simulations of a system
of $N$ particles interacting as
specified by  the Hamiltonian of
Eq.(\ref{en}).
 The procedure for computing the diffusion coefficient
goes as follows. The system is equilibrated
at a  fixed chemical potential and temperature.  In equilibrium
this system has $n$ particles.
Starting from this equilibrium configuration at a time $t=0$, each
one of these  $N$
particles is allowed to move to an empty neighbor site
randomly chosen. The move
is accepted  if the total energy of the system
is reduced by the move, otherwise it is accepted with
a probability $\exp(\Delta E /k_BT)$ where $\Delta E$ is
the difference between the energy of the system after and before the
move. After repeating this procedure  $Nt$ times, the mean
square displacement per particle  at a time $t$
is computed and the diffusion coefficient is obtained from
\begin{equation}
\overline{D}=\lim_{t\rightarrow \infty}\frac{\langle \Delta
\overline{r}(t)^2 \rangle}{6\overline{t}}\; .
\label{DD}
\end{equation}
where $\overline{r}=r/a$ and $a$ is the distance
between two neighbor sites and
$\overline{t}=t/t_{MC}$
is the time in Monte Carlo steps.

\section{Results and Conclusions}

\begin{figure}
\begin{center}
\includegraphics[clip=true,scale=0.6,width=6cm]{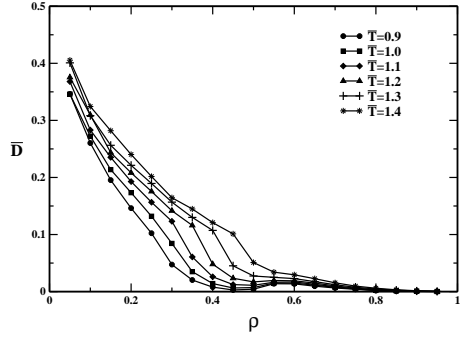}
\end{center}
\caption{Reduced Diffusion coefficient vs. density for $\overline{T}=0.9,1.1,1.2,1.3,1.4$.
$\overline{D}$ }
\label{cap:fig3}
\end{figure}

In order to find
if the three-dimensional associating lattice gas
exhibits diffusion anomalies, we
have analyzed how $\overline{D}$ varies with
the number density $\rho=N/(2L^3)$ for a fixed temperature.
Fig.  (\ref{cap:fig3})illustrates the behavior of $\overline{D}=D \tau_{MC}/a^2$ for $\overline{T}=k_BT/\varepsilon=0.9,1.1,1.2,1.3,1.4$ where $a$ is the lattice
distance and $\tau_{MC}$ is the tipical Monte Carlo time step. For
high temperatures, $\overline{T}>1.2$, the diffusion increases with decreasing 
density as in a normal liquid. 
\begin{figure}
\begin{center}
\includegraphics[clip=true,scale=0.6,width=6cm]{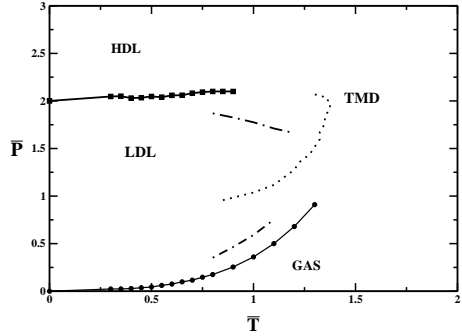}
\end{center}
\caption{Reduced pressure  vs. reduced temperature phase diagram showing the
the two liquid phase, two critical points, the density anomaly ( the TMD is the dotted line ) and the diffusion
 anomaly regions ( the temperature of maximum diffusion is the upper dot-dashed line and the temperature of minimum diffusivity is the lowe dot-dashed line).}
\label{cap:fig4}
\end{figure}

A different scenary appears for lower temperatures. 
The  reduced diffusion coefficient, $\overline{D}$ still decreases as $\rho$
increases for very low densitie. 
However, as the density is increased $\overline{D}$ has a minimum
at  $\rho_{Dmin}$, and  increases with
the increase of density from $\rho_{D_{min}}<\rho<\rho_{D_{max}}$.
Increasing the density above $\rho_{Dmax}$,  $\overline{D}$ decreases
again as expected. Therefore, there is a region
of densities $\rho_{Dmax}>\rho>\rho_{Dmin}$ where
the diffusion coefficient is anomalous, increasing
with density. This behavior is similar to the
diffusion anomaly present in SPC/E water.
A diffusion anomaly in the ALG model  is
observed in the range of temperatures $0.75<\overline{T}<1.2$
illustrated in Fig. (\ref{cap:fig4}).
The region in the $\overline{P}$ -$\overline{T}$ plane 
( where ($\overline{P}=P/\varepsilon/a^3$)   where there is an anomalous behavior in
 the diffusion is bounded by $(T_{D{\rm min}},P_{D{\rm min}})$ ( lower line )and
 $(T_{D{\rm max}},P_{D{\rm max}})$ (upper line) and 
lies partially inside the region of density anomalies 
what differs from the behavior observed  experimentally 
and in SPC/E water (see Fig. (\ref{cap:spce}) and Fig. (\ref{cap:experimental}))
but coincides with the behavior shown for non smooth ramp-like
potentials \cite{Ne06} that might be relevant for other tetrahedral 
materials.

In resume we have shown that the presence of a density anomaly
 seems to be associated with the
presence of diffusion anomaly, confirming observations
made in other models \cite{Ol06,Ku05} and in water
\cite{Ne01,Ne02b,Ne02}. This seems to indicate
that as the particles gain more energy by being
close together, this gain facilitates the mobility.
The hierarchy between the anomalies resembles the one observed in the
purely repulsive ramp-like discretized potential \cite{Ne06}. The link
between the two models is the presence of two competing
interaction distances and the non smooth transition between them.
The first ingredient seems to be the one that defines the presence
of the anomalies, while the second might govern the hierarchy between
them \cite{Ne06}\cite{Ol06}\cite{Ku05}. Similar behavior
should be expected in other models where the density anomaly is
also present \cite{Fr03}-\cite{Ca05}

\noindent{\Large\bf Acknowledgments}

We thank  the Brazilian science agencies CNPq, Capes, Finep and Fapergs
for financial support.

\bibliographystyle{elsart-num}
\bibliography{Biblioteca}

\end{document}